\documentclass[%
preprintnumbers,
nofootinbib,
 amsmath,amssymb,
 aps,
]{revtex4}

\usepackage{epsfig}
\usepackage{graphicx}% Include figure files
\usepackage{mathrsfs}
\usepackage{amsfonts}
\usepackage{caption2}
\usepackage{array}
\begin{document}

%\preprint{APS/123-QED}
\title{de Sitter Gauge Theory of Gravity: An Alternative Torsion Cosmology}

%\thanks{SRFDP under Grant No. 200931271104 and Shanghai Natural Science
%  Foundation, China Grant No. 10ZR1422000}%

\author{Xi-chen Ao}
\author{Xin-zhou Li}
\email{aoxichen@gmail.com, kychz@shnu.edu.cn}
\affiliation{Shanghai United Center of Astrophysics (SUCA),
Shanghai Normal University, 100 Guilin Road, Shanghai 200234, China}

\begin{abstract}
 A new cosmological model based on the de Sitter gauge theory (dSGT) is
 studied in this paper. By some transformations, we find, in the dust universe,
 the cosmological equations of dSGT could  form an autonomous system. We
conduct dynamics analysis to this system, and find 9 critical points,
among which there exist one positive attractor and one negative attractor. The
positive attractor shows us that our universe will enter a exponential expansion
phase in the end, which is similar to the conclusion of $\Lambda$CDM. We also
carry out some numerical calculations, which confirms the conclusion of dynamics
analysis. Finally, we fit the model parameter and initial values to the Union 2
SNIa dataset, present the confidence contour of parameters
and obtain the best-fit values of parameters of dSGT.
\end{abstract}

\keywords{torsion, cosmology beyond $\Lambda$CDM, gravity}
\maketitle

\section{Introduction}
\label{sec:introduction-2}
At the end of the last century, the astronomical observations of high redshift
type Ia supernovae (SNIa)  indicated that our universe is not only expanding, but also
accelerating, which conflicts with our deepest intuition of gravity. With some
other observations, such as cosmic microwave Background radiation (CMBR),
baryon acoustic oscillations (BAO) and large-scale structure (LSS), physicists
proposed a new standard cosmology model, $\Lambda$CDM, which introduces the
cosmological constant back again. Although this
unknown energy component accounts for 73\% of the energy density of the
universe, the measured value is too small to be explained by any current fundamental
theories.\cite{Peebles}-\cite{Hao}
If one tries to solve this trouble phenomenologically by setting the cosmological constant to a particular value,
the so-called fine-tuning problem would be brought up, which is considered as a basic problem almost any cosmological model would encounter.  A good model should restrict the
fine-tuning as much as possible. In order to alleviate this problem, various
alternative theories have been proposed and developed these years, such as
dynamical dark energy, modified gravity theories and even inhomogeneous universes.
Recently, a new attempt, called torsion cosmology, has attracted researchers' attention, which introduces dynamical torsion to mimic the contribution of the cosmological constant. It seems more natural to use a pure geometric quantity to account for the cosmic acceleration than to introduce an exotic energy component.

Torsion cosmology could be traced back to the 1970s, and the early work mainly
focused on issues of early universe, such as avoiding singularity and the origin
of inflation.  In some recent work, researchers attempted to extend the
investigation to the current evolution and found it might account for the cosmic acceleration. Among these models,
Poincar\'e gauge theory (PGT) cosmology is the one that has been
investigated most widely. This model is based on PGT, which is inspired by the
Einstein special relativity and the localization of global Poincar\'e
symmetry\cite{PGT}. Goenner \textit{ et al}. made a comprehensive survey of torsion
cosmology and developed the equations for all the PGT cases.\cite{Goenner}
Based on Goenner's work, Nester and his collaborators\cite{Nester}
found that the dynamical scalar torsion could be a possible reason for the
accelerating expansion. Li\textit{ et al}.\cite{Sun}
 extended the investigation to the late time evolution, which shows us the fate of our universe.

Besides PGT cosmology, there is another torsion cosmology, de Sitter gauge
theory (dSGT) cosmology, which can also be a possible explanation to the
accelerating expansion. This cosmological model is based on the de Sitter gauge
theory, in which gravity is introduced as a gauge field from de Sitter invariant special relativity
(dSSR), via the localization of de Sitter symmetry.\cite{Guo}
dSSR is a special relativity theory of the de Sitter space rather than the conventional Minkowski spacetime,
which is another maximally symmetric spacetime with an uniform scalar curvature
$1/R$. And the full symmetry
group of this space is de Sitter group, which unifies the Lorentz group and the
translation group, putting the spacetime symmetry in an alternatively
interesting way. But in the limit of $R\rightarrow \infty$, the de Sitter group
could also degenerate to the Poincar\'e group. To localize such a
global symmetry, de Sitter symmetry, requires us to introduce certain gauge
potentials which are found to represent the
gravitational interaction. The gauge potential for de Sitter gauge theory is the
de Sitter connecion, which combines Lorentz connection and orthonormal tetrad,
valued in $\mathfrak{s o}$(1,4) algebra. The gravitational action of dSGT takes
the form of Yang-Mills gauge theory. Via variation of the action with repect to
the the Lorentz connection and orthonormal tetrad, one could attain the
Einstein-like equations and gauge-like equations, respectively. These equations
comprise a set of complicated non-linear equations, which are difficult to
tackle. Nevertheless, if we apply them to the homogeneous and isotropic
universe, these equations would be much more simpler and tractable. Based on
these equations, one could construct an alternative cosmological model with
torsion. Analogous to PGT, dSGT has also been applied to the cosmology recently
to explain the accelerating expansion.\cite{Chaoguang}

Our main motivation of this paper is to investigate (i)whether the cosmological model based on
de Sitter gauge theory could explain the cosmic acceleration; (ii)where we are
going, i.e., what is the fate of our universe; (iii) the constraints of the
parameters of model imposed by means of the comparison of observational data. By some analytical and numerical
calculations, we found that, with a wide range of initial values, this model
could account for the current status of the universe, an accelerating expanding,
and the universe would enter an exponential expansion phase in the end.

This paper is organized as follows: First, we summarize the
de Sitter gauge theory briefly in Sec.~\ref{sec:de-sitter-gauge}, and then show the cosmological model
based on de Sitter gauge theory in Sec.~\ref{sec:cosm-evol-equat}. Second, we rewrite these dynamical
equations as an autonomous system and do some dynamical analysis and numerical
discussions on this system in the Sec.~\ref{sec:autonomous-system}
and~\ref{sec:numer-demonstr}.
Next in the \ref{sec:supern-data-fitt}th section, we compare the cosmological solutions to the SNIa data and constrain the
parameters. Last of all, we discuss and summarize the implications of our
findings in Section \ref{sec:summary-conclusion}.

\section{de Sitter Gauge Theory of Gravitation}
\label{sec:de-sitter-gauge}
\label{Supernovae Data Fitting}In dSGT, the de Sitter connection is introduced
as the gauge potential, which takes the form as
\begin{equation}
 (\check {B}^{AB}_{\ \ \ {\mu}})=\left(
\begin{array}{cc}
B^{ab}_{~~{\mu}} & R^{-1} e^a_\mu\\[0.1cm]
-R^{-1}e^b_\mu &0
\end{array}
\right ) \in \mathfrak{so}(1,4),
\end{equation}
where $\check{B}^{AB}_{\ \ \ \mu}=\eta^{BC}\check{B}^A_{~C\mu}$,
$\check{B}^{AB}_{\ \ \ 4}=\eta^{BC}\check{B}^A_{~C4}$
and $\eta^{AB}=\rm{diag}(1,-1,-1,-1,-1)$, which
combines the Lorentz connection and the  orthonormal tetrad \footnote{In this
  paper, the Greek indices, $\mu,\nu,...,$ are 4D coordinate indices, whereas the
  capital Latin indices $A,B,C,...,$  and the lowercase Latin indices,
  $a,b,...,$ denote 5D  and 4D orthonormal tetrad indices, respectively.} . The associated field strength is the curvature of
this connection, which is defined as
\begin{eqnarray}
  \label{eq:curvature}
  {\check {\cal F}}_{\mu\nu}= ( \check{\cal F}^{AB}_{~~~\mu\nu})
 =\left(
    \begin{array}{cc}
        F^{ab}_{~~\mu\nu} + R^{-2}e^{ab}_{~~ \mu\nu} & R^{-1} T^a_{~\mu\nu}\\[0.1cm]
        -R^{-1}T^b_{~\mu\nu} &0
    \end{array}
  \right )\in \mathfrak{so}(1,4),
\end{eqnarray}
where  $e^a_{~b\mu\nu}=e^a_\mu e_{b\nu}-e^a_\nu e_{b\mu},
e_{a\mu}=\eta_{ab}e^b_\mu$, $R$ is the de Sitter radius, and $ F^{ab}_{~~ \mu\nu}$ and $
T^a_{~\mu\nu}$ are the curvature and torsion
of Lorentz-connection,
\begin{eqnarray}
T^a_{~\mu\nu}&=&\partial_\mu
e^a_\nu-\partial_\nu e^a_ \mu+B^a_{~c \mu}e^c_\nu-B^a_{~c
\nu}e^c_\mu,\\
F^a_{~b
\mu\nu}&=&\partial_\mu B^a_{~b\nu} -\partial_ \nu
B^a_{~b\mu}+B^a_{~c\mu}B^c_{~b
\nu}-B^a_{~c\nu}B^c_{~b\mu},
\end{eqnarray}
which also satisfy the respective Bianchi identities.

The gauge-like action of gravitational fields in dSGT takes the form, \cite{Chaoguang}
\begin{eqnarray}
S_{\rm G}&=&\frac{\hbar}{4g^2}\int_{\cal M}d^4 x e
{\bf Tr}_{dS}(\check{\cal F}_{\mu\nu}\check{\cal F}^{\mu\nu})\\
&=& -\int_{\cal M}d^4 x e
\left[ \frac{\hbar}{4g^{2}} F^{ab}_{~~\mu\nu}F_{ab}^{~~\mu\nu}- \chi
 \left (F-\frac{6}{R^{2}}\right) - \frac{\chi}{2} T^a_{~\mu\nu}T_a^{~\mu\nu} \right].\label{GYM}
\end{eqnarray}
Here, $e=\det(e^a_\mu)$, $g$ is a dimensionless constant describing the
self-interaction of the gauge field, $\chi$ is a dimensional
coupling constant related to $g$ and $R$, and  $F=-\frac{1}{2}F^{ab}_{\
  \mu\nu}e_{ab}^{\ \mu\nu}$ is the scalar curvature of the Cartan connection. In
order to be consistent with Einstein-Cartan theory, we take
$\chi=1/(16\pi G)$ and $\hbar g^{-2}=3\chi \Lambda^{-1}$, where $\Lambda=3/R^{2}$.

Assuming that the matter is minimally coupled to gravitational fields, the total action of dSGT could be written as:
\begin{equation}\label{totalaction}
    S_T=S_{G}+S_M,
\end{equation}
where $S_{M}$ denotes the action of matter, namely the gravitational source.
Now we can obtain the field equations via variational principle with respect to $e^{a}_{\mu}, B^{ab}_{\ \ \mu}$,
\begin{eqnarray}\label{FEQ1}%
&&\nabla_{\nu}T_{a}^{~\mu\nu } -F_{~a}^\mu+\frac{1}{2}F e_a^\mu -\Lambda
e^{\mu}_{a}- \frac{8\pi G\hbar}{g^{2}}\left(e_{a}^\kappa {\rm Tr}(F^{\mu \lambda}F_{\kappa \lambda})-\frac{1}{4}e_a^\mu
{\rm Tr}(F^{\lambda \sigma} F_{\lambda \sigma})\right)\nonumber\\
&&  -16\pi G \chi\left(e_a^\kappa T_b^{~\mu\lambda}T^{b}_{~\kappa\lambda}-\frac{1}{4}e_a^\mu
T_b^{~\lambda\sigma}T^b_{~\lambda\sigma}\right)=8\pi
GT_{Ma}^{~\mu}\\[0.2cm]
\label{FEQ2}%
&&\nabla_{\nu}F_{ab}^{~~\mu\nu}-R^{-2}\left(Y^\mu_{~\,\lambda\nu}
e_{ab}^{~~\lambda\nu}+Y ^\nu_{~\, \lambda\nu } e_{ab}^{~~\mu\lambda}
+2T_{[a}^{~\mu\lambda} e_{b]\lambda}\right) = 16\pi GR^{-2}S^{\quad \mu}_{{\rm M}ab},%
\end{eqnarray}
where\begin{eqnarray}
T_{M a}^\mu :=-\frac{1}{e}\frac{\delta S_{M}}{\delta e^{a}_{\mu}}, \qquad
S_{M ab}^{\mu} :=\frac{1}{2\sqrt{-g}} \frac{\delta S_{M}}{\delta B^{ab}_{\mu}},
\end{eqnarray}
represent the effective energy-momentum density and spin density of the source,
respectively, and
\begin{eqnarray}
  Y ^\nu_{~\, \lambda\nu } := \frac{1}{2} (T^\lambda _{\ \,\nu\mu}+T^{\ \lambda}
  _{\mu \ \,\nu}+T^{\ \lambda} _{\nu \ \,\mu}),
\end{eqnarray}
is the contorsion. It is worth noticing that the Nabla operator in Eq. (\ref{FEQ1})
and (\ref{FEQ2}) is the covariant derivative compatible with Christoffel symbols \{$^{\mu}_{\nu\kappa}$\}
for coordinate indices, and  Lorentz connection $B_{b\mu}^a$ for
orthonormal tetrad indices. Readers can be referred to
Ref.\cite{Chaoguang}
for more details on dSGT.
\section{The Cosmological Evolution Equations}
\label{sec:cosm-evol-equat}
Since current observations favor a homogeneous, isotropic
universe, we here work on a Robertson-Walker (RW) cosmological metric
\begin{equation}
  \label{eq:RW}
  \mathrm{d}s^{2}=\mathrm{d}t^{2}-a (t)^{2} \left[ \frac{\mathrm{d}r^{2}}{1-kr^{2}}+r^{2}(\mathrm{d}\theta^{2}+\sin^{2}\mathrm{d}\phi^{2})\right].
\end{equation}
For Robertson-Walker metric, the nonvanishing torsion tensor components are of the form
\footnote{Here, the Latin indices i, j, k..., are 3D orthonormal tetrad indices
  with range 1, 2, 3.},
\begin{eqnarray}
  \label{eq:torsion-components}
\label{torsion}
  T^{i}_{j0}(t)=T_{+}(t)~\delta^{i}_{j}, \quad T^{i}_{jk}(t)=T_{-}(t)~\epsilon^{i}_{~jk},
\end{eqnarray}
where $T_{+}$ denotes the vector piece of torsion, namely, in components, the
trace of the torsion, and $T_{-}$ indicates the axial-vector
piece of torsion, which corresponds in components to the totally antisymmetric
part of torsion.
$T_{+}$ and $T_{-}$ are both functions of time $t$, and their subscripts, +
and -, denote the even and odd parities, respectively.

The nonvanishing torsion 2-forms in this case are
\begin{eqnarray}
{\bf T}^0 &=& 0 \nonumber \\
{\bf T}^1 &=& {T_+}\,  {\vartheta}^0\wedge {\vartheta}^1 +
{T_-}\,  {\vartheta}^2\wedge {\vartheta}^3\nonumber \\
{\bf T}^2 &=&  {T_+}\,  {\vartheta}^0\wedge {\vartheta}^2 +  {T_-}\,  {\vartheta}^3\wedge {\vartheta}^1  \\
{\bf T}^3 &=& {T_+}\,  {\vartheta}^0\wedge {\vartheta}^3 +  {T_-}\,  {\vartheta}^1\wedge {\vartheta}^2 , \nonumber
\end{eqnarray}
where $\vartheta^{0}=\mathrm{d} t,\ \vartheta^{1}=\frac{a(t)\mathrm{d} r}{\sqrt{1-k r^2}},\ \vartheta^{2}=a(t)r\mathrm{d}   \theta\
\mathrm{and} ~ \vartheta^{3}=a(t)r\sin\theta \mathrm{d} \phi $.

According to the RW metric Eq.\eqref{eq:RW} and the torsion Eq.~\eqref{torsion},
the field equations could be reduced to
 \begin{eqnarray}
\label{el-00}%
&& - \frac {\ddot a^2} {a^2}
 -  \left(\dot T_++ 2\frac{\dot a}{ a} T_+ -2\frac {\ddot a}{a} \right)\dot T_+
 + \frac1 4 \left(\dot T_-+2\frac{\dot a} aT_- \right)\dot T_-
+ T_+^4-\frac {3}{2}
T_+^2 T_-^2+ \frac{1} {16} T_-^4 + \left(5 \frac{\dot a^2} {a^2}\right. \nonumber \\
&& \quad \left. + 2\frac{ k} {a^2}-\frac{3}{R^2}\right) T_+^2-\frac{1}{2}
\left(\frac{5}{2}\frac{\dot a^2} {a^2} + \frac{k}{a^2}  -\frac{3}{R^2}\right) T_-^2 + 2 \frac
{\dot a} a \left(\frac{\ddot a} {a}  - 2 \frac{\dot a^2} {a^2}-2\frac{ k} {a^2}
+\frac{3}{R^2}\right)T_+ - \frac{\dot a} {a} (4  T_+^2 \nonumber\\
&&\quad  - 3  T_-^2)T_+ +\frac{\dot a^2}{a^2}\left( \frac
{\dot a^2}{a^2}  + 2 \frac{k} {a^2}- \frac{2}{R^2}\right) +\frac{k^2}{a^4}  - \frac{2}
{R^2} \frac{k}{a^2} +\frac{ 2}{R^4}=-\frac{16\pi G\rho}{3 R^2},
\\[0.3cm]
 \label{el-11}%
&&\frac{\ddot a^2} {a^2} + \left(\dot T_+ + 2\frac{\dot a} a  T_+ - 2\frac{\ddot a} a
+ \frac{6}{R^2}\right)\dot T_+ -\frac 1 4 \left(\dot T_-
+ 2 \frac {\dot a} a T_-\right)\dot T_- - T_+^4 + \frac 3 2 T_+^2 T_-^2 - \frac1 {16} T_-^4 \nonumber\\
&&\quad+ \frac {\dot a} a(4  T_+^2 - 3 T_-^2)T_+  - \left(5\frac{\dot a^2} {a^2} + 2 \frac k
{a^2}  + \frac3 {R^2}\right) T_+^2+ \frac 1 2 \left(\frac 5 2\frac{\dot a^2} {a^2} +
\frac k {a^2} +  \frac 3 {R^2}\right) T_-^2- 2\frac{\dot a} a \left(\frac{\ddot a } {a}- 2\frac{\dot a^2} {a^2 }\right.\nonumber \\[0.2cm]
&&\quad \left. - 2  \frac k {a^2}- \frac6 {R^2}\right)T_+  - \frac 4 {R^2} \frac{\ddot a} a -\frac{\dot a^2} {a^2} \left(\frac{\dot a^2}{a^2} +2\frac k
{a^2}\right )+  \frac2 {R^2}
-\frac{k^2}{a^4}  - \frac2 {R^2}\frac k {a^2} +\frac6 {R^4}
= -\frac{16\pi G p}{R^2},
\\[0.3cm]
\label{yang1} %
&&\ddot T_-  + 3 \frac{\dot a} a \dot T_- + \left( \frac 1 2 T_-^2 - 6 T_+^2 + 12 \frac {\dot a} a
T_+  +\frac{\ddot a} a - 5\frac{\dot a^2}{a^2}
-  2\frac k {a^2}+  \frac 6 {R^2}\right)  T_-=0,
\\[0.3cm]
\label{yang2}%
&&  \ddot T_+ + 3 \frac{\dot a} a \dot T_+ -\left( 2  T_+^2  -\frac 3 2 T_-^2 - 6\frac{\dot a} a
T_+ -\frac {\ddot a} a  + 5 \frac {\dot a^2}  {a^2} + 2 \frac k {a^2}- \frac 3 {R^2}\right)
T_+ - \frac 3 2 \frac{\dot a} a T_-^2-\frac{\dddot a} a - \frac{\dot a\ddot a}{a^2}
\nonumber\\
&& \quad+ 2\frac {\dot a^3} {a^3} + 2\frac{\dot a} a \frac k {a^2} =0, %
\end{eqnarray}
where Eqs.\eqref{el-00} and \eqref{el-11} are the $(t,t)$ and $(r,r)$ component of
Einstein-like equations, respectively; and Eqs.\eqref{yang1} and \eqref{yang2} are
2 independent Yang-like equations, which is derived from the
$(r,\theta,\phi)$ and $(t,r,r)$ components of Lorentz connection.
The spin density of present time is generally thought be
very small which  could be neglected. Therefore, we here assumed the spin density is zero.

The Bianchi identities ensure that the energy momentum tensor is conserved,
which leads to the continuity equation:
\begin{equation}
  \label{eq:continuity}
  \dot{\rho}=-\frac{3\dot{a}}{a}(\rho+p).
\end{equation}
Equation \eqref{eq:continuity} can also be derived from
Eqs.~\eqref{el-00}-~\eqref{yang2}, which means only four of
Eqs.~\eqref{el-00}-~\eqref{eq:continuity} are independent.
With the equation of state (EoS) of matter content, these four equations comprise a complete system
of equations for five variables, $a(t),T_+(t),T_-(t),\rho(t)~\mathrm{and }~p(t)$.
By some algebra and differential calculations, we could simplify these 5 equations as:
 \begin{eqnarray}
\label{eq:Hubble}
\dot{H}&=&-2H^{2}-\frac{k}{a^{2}}+\frac{2}{R^{2}}+\frac{4\pi
  G}{3}(\rho+3p)+\frac{3}{2}\left(\dot{T}_++3H
  T_{+}-T_{+}^{2}+\frac{T_-^2}{2}\right)
\nonumber\\ &&+(1+3w)\rho,\\
\ddot{T}_+&=&-3\left(H +\frac 3 2 T_+\right)\dot{T}_+ -3T_{-}\dot{T}_- -\frac{8\pi G}{3}(\rho+3p)^.
-\frac{3}{2}H T_{-}^{2}+\left[\frac {13} 2 ({T_+}-3 H){T_+}\right.\nonumber \\
&&
\left.+ 6H^2+\frac{3k}{a^2}+\frac{9T_-^2}{4}-\frac 8 {R^2}-\frac{28\pi G}{3}(\rho+3p)\right]T_+,\\
\ddot{T}_- &=&-3H\dot{T}_- -\left[-\frac{15}{2}T_{+}^{2}+\frac{33H T_{+}}{2}-6H^{2}-\frac{3k}{a^2}+\frac{8}{R^{2}}+\frac{5}{4}T^{2}_{-}+\frac{3}{2}\dot{T}_+ \right.\nonumber\\
&&\left. + \frac{4\pi G}{3}(\rho+3p)\right]T_{-},\\
\dot{\rho}&=&-3H(\rho+p),\\
\label{eq:EoS}
w&=&\frac{p}{\rho},
 \end{eqnarray}
where $H=\dot{a}/a$ is the Hubble parameter.

\section{Autonomous System}
\label{sec:autonomous-system}
If we  rescale the variables and parameters as
\begin{eqnarray}
&&t\rightarrow t/l_0;\quad H\rightarrow l_0 H;\quad k\rightarrow l_0^{2}k;\quad R\rightarrow R/l_0;\nonumber \\
&&T_{\pm}\rightarrow l_0 T_{\pm};\quad \rho \rightarrow \frac{4\pi G l_{0}^2}{3 }\rho;\quad p \rightarrow \frac{4\pi G l_{0}^2}{3 }p,\label{transformation}
\end{eqnarray}
where $l_0=1/H_0$ is the Hubble radius in natural units, these variables and parameters would be
dimensionless. Under this transformation, Eqs.~\eqref{eq:Hubble}-\eqref{eq:EoS} remain unchanged expect for
the terms including $4\pi G\rho/3$ and $4\pi G  p/3$, which change into $\rho$ and $p$ respectively.
The contribution of radiation and spatial curvature in current universe
are so small that it could be neglected, so we here
just consider the dust universe with spatial flatness, whose
EoS is equal to zero.
By some further calculations, these equations could be transformed to a
set of six one-order ordinary derivative equations, which forms a
six-dimensional autonomous system, as follows,
\begin{eqnarray}\label{dia}
\dot{H}&=&-2H^{2}+\frac{2}{R^{2}}+\frac{3}{2}\left(P+3H T_{+}-T_{+}^{2}+\frac{T_-^2}{2}\right)+\rho,\\
\dot P &=&
-3\left(H +\frac 3 2 T_+\right)P -3T_{-}Q-\frac{3}{2}H T_{-}^{2}+ \left[\frac {13} 2 ({T_+}-3 H){T_+}+ 6H^2\right.
\nonumber \\ && \left.+\frac{9T_{-}^{2}}{4}-\frac 8 {R^2}
- 7\rho\right]T_+ + 6H\rho ,\\[0.2cm]
\dot{T_{+}}&=&P,\\
\dot Q&=&-3H Q -\left(-\frac{15}{2}T_{+}^{2}+\frac{33H
    T_{+}}{2}-6H^{2}+\frac{8}{R^{2}}+\frac{5}{4}T^{2}_{-}+\frac{3}{2}P + \rho \right)T_{-},\\
\dot{T_{-}}&=&Q,\\
\dot{\rho}&=&-3H\rho. \label{rho}
\end{eqnarray}
For such an autonomous system, we can use the dynamics analysis to
investigate its qualitative properties. Critical points are some exact constant
solutions in the autonomous system, which indicate the asymptotic behavior
of evolution. For example, some solutions, such as heteroclinic orbits, connect
two different critical points, and some others, such as homoclinic orbits, are a closed loop starting from and
coming back to the same critical point. In the dynamics analysis of cosmology,
the heteroclinic orbit is more interesting.\cite{Zhao}
Thus, critical points could be treated as the basic tool in dynamics analysis, form which one could know the qualitative  properties
of the autonomous system.
By some algebra calculation, we find all the 9 critical points ($H_{c},\ P_{c},\
 Q_{c},\ T_{+c},  \ T_{-c}, \rho_{c}$) of this system, as shown in Table 1.
\begin{table}[t]
\centering
\begin{tabular}{| p{0.8cm} p{3.5cm} p{7.2cm} | }\hline
   &Critical Points  & Eigenvalues \\
\hline
&&\\[-0.3cm]
(i)&$(\frac{1}{R},0,0,0,0,0)$&
$-\frac{1}{R},-\frac{1}{R},-\frac{2}{R},-\frac{2}{R},-\frac{3}{R},-\frac{4}{R}$
\\[0.12cm]
(ii)&$(-\frac{1}{R},0,0,0,0,0)$&
$\frac{1}{R},\frac{1}{R},\frac{2}{R},\frac{2}{R},\frac{3}{R},\frac{4}{R}$  \\[0.12cm]
(iii)&$(-\frac{1}{2R},0,0,-\frac{2}{R},0,0)$&
$-\frac{2}{R},\frac{2}{R},\frac{3}{2R},-\frac{5}{2R},\frac{7}{2 R},\frac{4}{R}$
\\[0.12cm]
(iv)&$(\frac{1}{2R},0,0,\frac{2}{R},0,0)$&
$-\frac{2}{R},\frac{2}{R},\frac{5}{2R},-\frac{3}{2R},-\frac{4}{R},-\frac{7}{2R}$
\\[0.12cm]
(v)&$(-\frac{1}{2R},0,0,\frac{1}{2R},0,0)$&
$\frac{1}{2R},\frac{1}{R},-\frac{1}{R},\frac{2}{R},\frac{3}{2R},\frac{5}{2R}$ \\[0.12cm]
(vi)&$(\frac{1}{2R},0,0,-\frac{1}{2R},0,0)$&
$-\frac{1}{2R},\frac{1}{R},-\frac{1}{R},-\frac{3}{2R},-\frac{5}{2R},-\frac{2}{R}$
 \\[0.12cm]
(vii)&$(0,0,0,-\frac{\sqrt{3/2}}{R},0,\frac{1}{4R^{2}})$&$-\frac{\sqrt{3}}{\sqrt{2}R},
\frac{\sqrt{3}}{\sqrt{2}R}, -\frac{\sqrt{3}}{R}, \frac{\sqrt{3}}{R}, -\frac{\sqrt{6}}{R}, \frac{\sqrt{6}}{R} $
 \\[0.12cm]
(viii)&$(0,0,0,\frac{\sqrt{3/2}}{R},0,\frac{1}{4R^{2}})$&$-\frac{\sqrt{3}}{\sqrt{2}R},
\frac{\sqrt{3}}{\sqrt{2}R}, -\frac{\sqrt{3}}{R}, \frac{\sqrt{3}}{R}, -\frac{\sqrt{6}}{R}, \frac{\sqrt{6}}{R} $ \\[0.2cm]
(ix)&$(0,0,0,0,0,\frac{-2}{R^{2}})$  & $-\frac{\sqrt{3}-3\rm{i}}{\sqrt{2}R},
\frac{\sqrt{3}-3\rm{i}}{\sqrt{2}R}, -\frac{\sqrt{3}+3\rm{i}}{\sqrt{2}R},
\frac{\sqrt{3}+3\rm{i}}{\sqrt{2}R}, -\frac{\rm{i}\sqrt{6}}{R}, \frac{\rm{i}\sqrt{6}}{R}  $     \\[0.10cm]
\hline
\end{tabular}
\caption{  \label{tab:critical-points}The critical points and their corresponding eigenvalues. The point 9
  is not physically acceptable, for its negative energy density.}
\end{table}
Furthermore, we analyze the stabilities of these critical points by means of the
first-order perturbations. Substituting these linear perturbations into these
dynamcial equations, we would obtain the perturbation equations around the
critical points, i.e.
\begin{align}
\delta \boldsymbol{\dot{x}} = A \boldsymbol{x}, &  \quad A = \frac{\partial \boldsymbol{f}}{\partial \boldsymbol {x}}|_{\boldsymbol{x}=\boldsymbol{x}_c},
\end{align}
where $x$ means the six variables of this autonomous system and $f$ denotes the
corresponding vector function on the right-hand side of
Eqs.~\eqref{dia}-\eqref{rho}. Using the coefficient matrix $A$'s eigenvalue, we could analyze the
stabilities of these critical points. And the classification of these critical points
is shown in Table~\ref{tab:stabilities}. Among these critical points, there are only one  positive attractor, i.e. point (i),
whose eigenvalues are all negative, and only one negative attractor, i.e. point
(ii), whose eigenvalues are all positive. The negative attractor works as a source, from which
the phase orbits start off, whereas the positive attractor works as a sink, which the orbits finally approach.
And it is the heteroclinic line that connects
the positive attractor and the negative attractor, as shown in Fig. 1. Positive attractors are stable exact
solutions, describing the infinite future behavior of evolution, while the unstable negative attractors
depict the stories of infinite past. Therefore the positive attractor, point (i),
here shows us the picture of late time universe, where all quantities tend to
zero, except the Hubble parameter which approaches a finite value. At that time, the whole
universe is entering an exponential expansion phase, just like the $\Lambda$CDM
model.
\begin{table}
\centering
\begin{tabular}{|p{2.5cm}p{3.5cm}p{1.5cm}|}\hline
 Critical Points&  Property&  Stability \\
\hline
 (i)& Positive-Attractor&  Stable\\[0.05cm]
 (ii)&  Negative-Attractor&  Unstable\\[0.05cm]
 (iii)&  Saddle & Unstable\\[0.05cm]
 (iv)&  Saddle & Unstable\\[0.05cm]
 (v)&  Saddle & Unstable\\[0.05cm]
 (vi)&  Saddle & Unstable\\[0.05cm]
 (vii)&  Saddle&   Unstable\\[0.05cm]
 (viii)&  Saddle& Unstable\\[0.05cm]
 (ix)&  Spiral-Saddle & Unstable\\[0.05cm]
\hline
\end{tabular}
\caption{  \label{tab:stabilities}The stability properties of critical points}
\end{table}
\begin{figure}
\centering
\includegraphics[width=8cm,height=6.5cm]{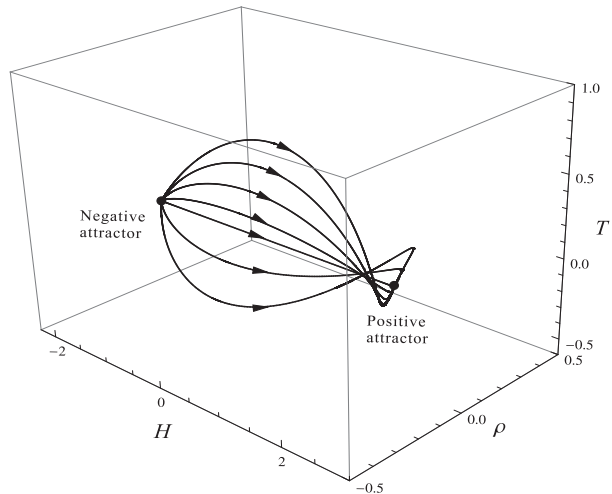}
\caption{ \label{fig:heteroclinic}The $(H,T_+,\rho)$ section of the phase diagram with R=4/3. The heteroclinic orbits connect the critical point (i) and (ii).}
\end{figure}

\section{Numerical Demonstration}
\label{sec:numer-demonstr}
In order to confirm these qualitative results derived from dynamics analysis
and know better about the global properties of this model, we explore the autonomous
system  by numerical methods. To solve the Eqs.~\eqref{dia}-\eqref{rho}
numerically, we choose some generic initial conditions and
parameters, as shown in Table \ref{tab:numerical-demonstrations}.
\begin{table}[h]
\centering
\begin{tabular}{|p{1.2cm}  p{1.cm} p{1.cm}p{1.cm}p{1.cm}p{1.cm}p{1.cm}p{0.5cm}|}\hline
 Case& $R$ &$H_{0}$ &$P_{0}$ &$Q_{0}$&$T_{+0}$&$T_{-0}$ &$\rho_{0}$\\
\hline
&&&&&&&\\[-0.25cm]
 (a.1)& 1.5 &1&0 &0&0&0&0.5\\
 (a.2)& 1.5 &1&0 &-0.5&-0.5&0&1\\
 (a.3)& 1.5 &1&-0.75 &-1&2&1.2&0.7\\
\hline
&&&&&&&\\[-0.25cm]
 (b.1)&0.4 &1&0 &0&-1.5&0&0.8\\
 (b.2)&0.6 &1&0 &0&-1&0&1\\
 (b.3)&1.5 &1&0 &0&0&0&0.5\\
\hline
\end{tabular}
\caption{  \label{tab:numerical-demonstrations}The values of initial conditions and
  parameters for the evolution curves in Fig.~\ref{fig:vtvr}. }
\end{table}
 First, we vary initial conditions $(P_{0} ,
Q_{0},T_{-0},T_{+0},\rho_{0})$ with a fixed de Sitter radius, and the results are shown in Fig.2(a).
Then we change the de Sitter radius, and show the results in Fig.2(b). Because of the rescaling Eqs.~\eqref{transformation},
the current Hubble parameter here must be 1.
From these numerical results, it is easy to find that the
Hubble paramter of all the solutions approaches a particular finite value in the infinite future, whatever
the initial conditions are, and this value only depends on the de Sitter
radius $R$. Such results demonstrate the dynamics analysis we have done in the former section.
We could also find that this positive attractor covers a wide range of initial conditions, and therefore the
troublesome fine-tuning problem has been alleviated.  In comparison with the result of PGT, we
find  the cosmology based on de Sitter gauge theory is quite different from the Poincare gauge theory, where the
expansion will asymptotically come to a halt.
It is the existence of the de
Sitter radius that makes such a discrepancy. If we let $R\rightarrow\infty$, the
de Sitter gauge theory would degenerate to the PGT.
\begin{figure}[t]
\centering
\includegraphics[width=7.2cm,height=5.5cm]{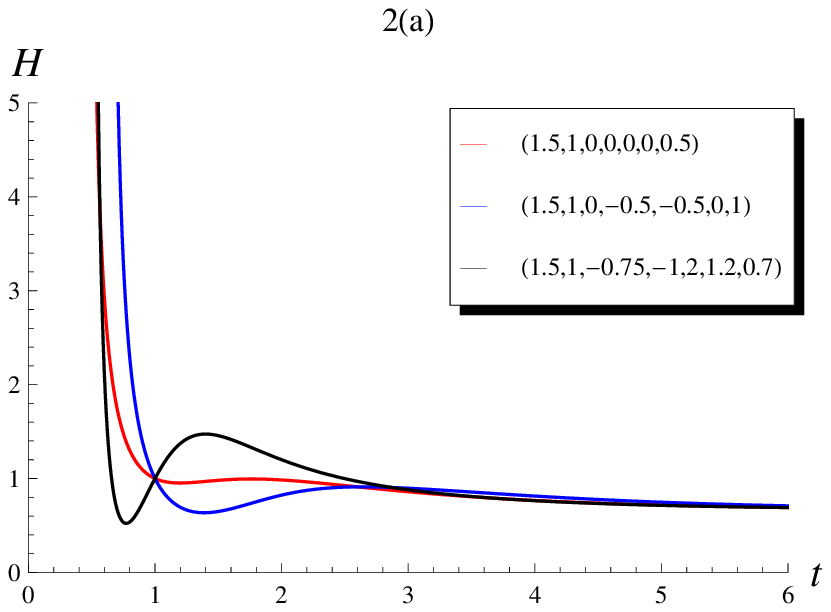}
\includegraphics[width=7.2cm,height=5.5cm]{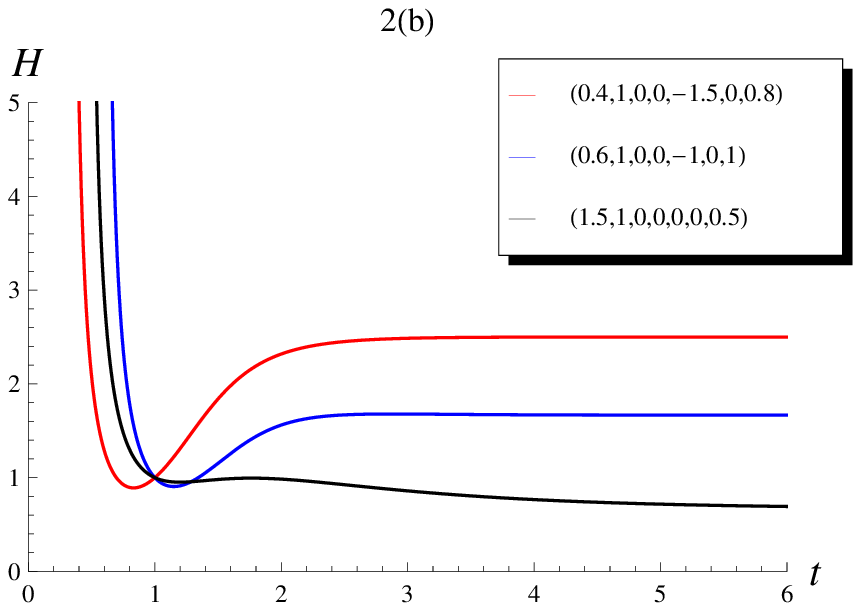}
\caption{ \label{fig:vtvr}The evolution of Hubble parameter $H$ with respect to some initial values and parameter choice $(R,H_0,P_0,Q_0,T_{+0},$ $T_{-0},\rho_0)$. According to the transformations (\ref{transformation}), the unit of time here is the Hubble Time. In Fig 2(a), we fixed $R$ and changed $T_\pm$, while in Fig 2(b), we changed $R$.}
\end{figure}
\section{Supernovae Data Fitting}
\label{sec:supern-data-fitt}
A basic  approach to testing a cosmological model is the supernova fitting
through its description of the expansion history of the universe. In this
section we fit the initial conditions and model parameters to current type Ia supernovae data.
And the maximum likelihood technique is used here, which could determine the best fit values
of parameters and initial conditions and the goodness of this model.
The supernova data are comprised of the distance modulus $\mu_{obs}$, which
is equal to the difference between apparent magnitude $m_{i}$  and absolute
magnitude $M_{i}$, and redshifts $z_{i}$ of supernovae with their corresponding
errors $\sigma_{i}$. Note that the error here are assumed to be normally distributed and independent.

The theoretical distance modulus is related to the luminosity distance $d_{L}$ by
\begin{eqnarray}
  \label{eq:modulus}
  \mu_{th}(z_{i})&=&5 \log_{10}\left(\frac{d_{L}(z_{i})}{\mathrm{Mpc}}\right)+25 \nonumber\\
&=&5 \log_{10}D_{L}(z_{i})-5\log_{10}\left( \frac{c H_{0}^{-1}}{\mathrm{Mpc}} \right)+25 \nonumber\\[0.12cm]
&=& 5 \log_{10}D_{L}(z_{i})-5\log_{10}h +42.38,
\end{eqnarray}
where the $D_{L}(z)$ is the dimensionless 'Hubble-constant free' luminosity defined
by $D_{L}(z)=H_{0}d_{L}(z)/c$.

For a spatially flat cosmological model, which we consider here, the luminosity
distance could be expressed in terms of Hubble parameter  $H(z)$, as follows,
\begin{eqnarray}
  \label{eq:dL}
D_{L}(z)&=& (1+z)\int^{z}_{0} \mathrm{d}z' \frac{1}{H(z'; a_{1},...a_{n})},
\end{eqnarray}
where the Hubble parameter $H(z'; a_{1},...a_{n})$ here is the dimensionless Hubble parameter under the
rescaling transformation Eq \eqref{transformation}.

As we know, due to the normal distribution of errors, we could use the
$\chi^{2}$ parameter as the maximum likelyhood estimator to determine the best fit values of
parameters and initial conditions ($R,P_{0},Q_{0},T_{+0},T_{-0},\rho_0$) of the model.
The $\chi^{2}$ here for the SNIa data is
\begin{eqnarray}
  \label{eq:chi2}
  \chi^{2}(\theta)&=&\sum^{N}_{i} \frac{[\mu_{obs}(z_{i})-\mu_{th}(z_{i})]^{2}}{\sigma_{i}^{2}},\nonumber\\
&=&\sum^{N}_{i} \frac{[\mu_{obs}(z_{i})-5\log_{10}D_{L}(z_{i};\theta)-\mu_{0}]^{2}}{\sigma_{i}^{2}},
\end{eqnarray}
where $\mu_{0}=-5\log_{10}h + 42.38$ , $\theta$ denotes all the parameters and
initial conditions, and $\sigma_{i}$ are the statistical errors
of SNIa.  If we want to include the systematic errors, which are comparable to
the statistical errors and should be taken into account seriously, we could
resort to the covariance matrix $C_{SN}$, and the Eq.~\eqref{eq:chi2} turn out
to take the form
\begin{eqnarray}
  \label{eq:chi22}
 \chi^{2}(\theta)&=&\sum^{N}_{i,j}\left[\mu_{obs}(z_{i})-\mu_{th}(z_{i})\right]
 (C_{SN}^{-1})_{i j}[\mu_{obs}(z_{j})-\mu_{th}(z_{j})],\\
&=&\sum^{N}_{i,j}\left[\mu_{obs}(z_{i})-5\log_{10}D_{L}(z_{i};\theta)-\mu_{0}\right]
 (C_{SN}^{-1})_{i j}[\mu_{obs}(z_{j})-5\log_{10}D_{L}(z_{i};\theta)-\mu_{0}].
\end{eqnarray}
The parameter $\mu_{0}$ here is a nuisance parameter, whose contribution we
are not interested in. So we marginalize over this parameter, $\mu_{0}$, thus
obtaining a new $\chi^2$,
\begin{eqnarray}
 \label{eq:chi23}
\chi^{2}(\theta)=A(\theta)-\frac{B(\theta)^{2}}{C}+\ln\left(\frac{C}{2\pi}\right),
\end{eqnarray}
where
\begin{eqnarray}
 \label{eq:marginalization}
A(\theta)&=&\sum^{N}_{i,j}\left[\mu_{obs}(z_{i};\theta)-5\log_{10}D_{L}(z_{i};\theta)\right]
 (C_{SN}^{-1})_{i
   j}[\mu_{obs}(z_{j};\theta)-5\log_{10}D_{L}(z_{j};\theta)],\\
B(\theta)&=&\sum^{N}_{j} (C_{SN}^{-1})_{i   j}[\mu_{obs}(z_{j};\theta)-5\log_{10}D_{L}(z_{j};\theta)],\\
C(\theta)&=&\sum^{N}_{i,j}=(C_{SN}^{-1})_{ij}.
\end{eqnarray}

Now we try to constrain our model parameter and initial values by this maximum
likelihood estimator. The dataset we use here is the ``Union2'' SNIa
dataset (N=557), the most comprehensive one to date, which combines all the
previous SNIa dataset in a homogeous manner.

\begin{figure}[t]
\centering
\includegraphics[width=17cm,height=9cm]{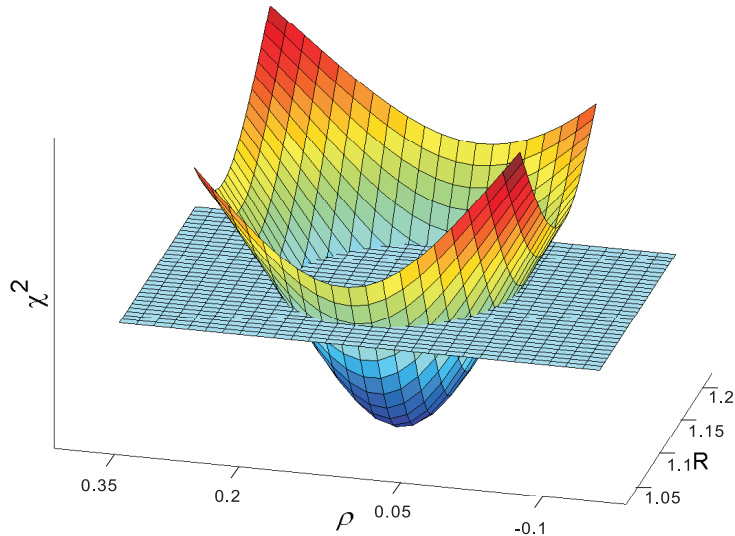}
\caption{ \label{fig:chi2-distribution}The $\chi^{2}$ distribution with respect to $R$ and $\rho$, compared
  to the $\Lambda$CDM, the plane $\chi^{2}=536.634$. Here we assume that all the torsions and their first
order derivatives vanish at present time.}
\end{figure}

\begin{figure}
\centering
\includegraphics[width=15.cm,height=7cm]{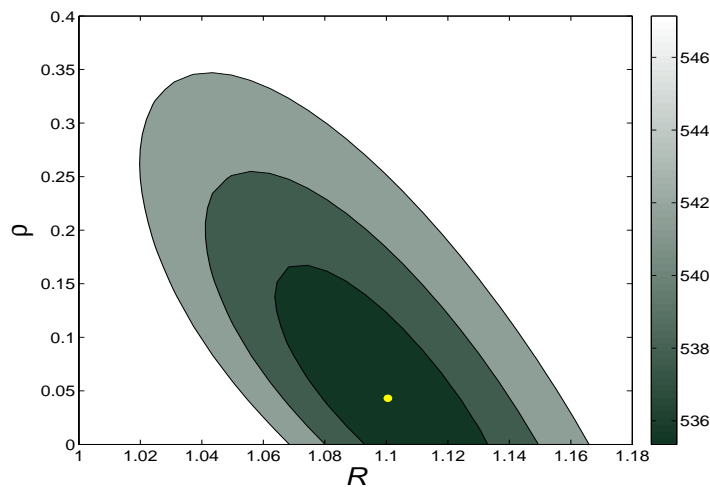}
\caption{ \label{fig:contour}The 68.3\%, 95.4\% and
  99.7\% $\chi^{2}$ confidence contours of dSGT with respect to $R$ and $\rho$, using the Union 2 dataset. Here we also
  assume that the current-time values for  all the torsions and their first
  order derivatives are zero. The yellow point is the best-fit point.  }
\end{figure}

By minimizing the $\chi^{2}$, we find the best fit parameters of dSGT model, as
shown in Table.~\ref{tab:best-fit}
\begin{table}[h]
\centering
\begin{tabular}{|p{0.8cm}p{1cm} p{1cm} p{1.cm}p{1.cm}p{1.cm}p{1.cm}p{1.2cm}|}\hline
Case& $R$  &$P_{0}$ &$Q_{0}$&$T_{+0}$&$T_{-0}$ &$\rho_{0}$&$\chi^{2}$\\
\hline
&&&&&&&\\[-0.25cm]
I. & 1.1005 & 0   &  0   &  0  & 0 & 0.0431 &535.3384\\
\hline
\end{tabular}
\caption{\label{tab:best-fit}The best-fit  initial data and parameters }
\end{table}
Based on the current observations, the present density of torsion in our universe is
very small, so it is reasonable to assume that the initial values of all torsions
and their first order derivative are zero at $z=0$. But their second order derivatives
does not vanish yet, which would have a significant impact on the history and
future of the evolution of our universe. In this case, the number of  parameter and initial
value is reduced, and the rest parameters and initial values are just $R$ and
$\rho_{0}$. It is easy to find the best-fit of these 2 parameters, which are shown in
Table \ref{tab:best-fit}.  And the minimal $\chi^{2}$ is $535.3384$,
whereas the value for $\Lambda$CDM is $536.634$, with $\Omega_{m}=0.27,
\Omega_{\Lambda}=0.73$.
In Fig.~\ref{fig:chi2-distribution}
we show the $\chi^{2}$ distribution with
respect to $R$ and $\rho$ compared to $\Lambda$CDM model, where the plane
$\chi^{2}=536.635$ corresponds to the value of $\Lambda$CDM.  Furthermore, we plot the contours of some
particular confidence levels, as shown in Fig.~\ref{fig:contour}. From these
figures, we could find that the evolution of our universe is insensitive to the initial value, which alleviate the fine-tuning problem.

\section{Summary and Conclusion}
\label{sec:summary-conclusion}
The astronomical observations imply that our universe is accelerating to a de
Sitter spacetime. This gives us a strong motive to consider the cosmic evolution
based on the  de Sitter gauge theory instead of other gravity theories.  The
localization of de Sitter symmetry requires us to introduce curvature and
torsion. So in de Sitter gauge theory, the torsion is an indispensable
quantity, by which people tried to include the effect of spin density in gravity
theory at first. But now this essential quantity might account for the acceleration
of our universe, if we apply dSGT to cosmology.

We found the cosmological equations for
dust universe in dSGT could form an autonomous system by some transformations,
where the evolution of the universe is described in terms of the orbits in phase space.
Therefore, by dynamics analysis to the dSGT cosmology, one could study the
qualitative properties of this phase space. We found all 9 critical points,
as shown in Table~\ref{tab:critical-points}. We also analyzed the
stabilities of these critical points, and found among these critical points there is only one positive attractor, which is stable.
The positive attractor alleviates the fine-tuning problem and  implies that the universe
will expand exponentially in the end, whereas all other physical quantities will turn
out to vanish. In this sense, dSGT cosmology looks more like the $\Lambda$CDM,than
PGT cosmology. And we conducted some concrete numerical calculations of this
the destiny of our model of the universe, which confirms conclusions from dynamics analysis.

Finally, in order to find the best-fit values and constraints of model parameters and initial conditions,
we fitted them to the Union 2 SNIa dataset. The maximum
likelihood estimator here we used is the $\chi^{2}$ estimate. By
minimizing the $\chi^{2}$, we found the best-fit parameters $R=1.135,~\rho=0.274$ and the
corresponding $\chi^{2}=535.3384$, while the value for $\Lambda$CDM is $536.634$,
with $\Omega_{m}=0.27, \Omega_{\Lambda}=0.73$. Note that we here set all the
initial values of torsions and their first-order derivatives to zero at $t=t_{0}$, since the
contribution of torsion to the current universe is almost negligible. We also plotted the confidence
contour Fig.~\ref{fig:contour} with respect to $R$ and $\rho$, from which it is easy to see that the fine-tuning problem
is alleviated and the evolution is not so sensitive to the initial values and model
parameters.

If we want to go deeper into cosmology based on de Sitter gauge theory, there
are a  lot of work need to be
done. We should fit this model to some other observations, like BAO and LSS
etc, to constrain the parameters better.  We also could study the perturbations
in the early universe, and compare the results to CMBR data. These issues will
considered in some upcoming papers.

\section*{Acknowledgments}This work is supported by SRFDP under Grant
No. 200931271104 and Shanghai Natural Science Foundation, China, under  Grant No. 10ZR1422000.

\end{document}